\documentclass[%
 reprint,
 longbibliography,
 amsmath, amssymb,
 aps, prl,
superscriptaddress
]{revtex4-2}

\usepackage[version=4]{mhchem}
\usepackage{graphicx}
\usepackage{dcolumn}
\usepackage{bm}
\usepackage{physics}
\usepackage{romannum}
\usepackage{hyperref}
\usepackage{color}
\def\new{\color{black}}

\begin{document}


\title{Sizable Ligand-Mediated Bond-Dependent Interactions in a Spin-1 Triangular Antiferromagnet \ce{NiI2}}

\affiliation{National Laboratory of Solid State Microstructures and School of Physics, Nanjing University, Nanjing 210093, China} %
\affiliation{Key Laboratory of Computational Physical Sciences (Ministry of Education), Institute of Computational Physical Sciences, State Key Laboratory of Surface Physics, and Department of Physics, Fudan University, Shanghai 200433, China} %
\affiliation{Collaborative Innovation Center of Advanced Microstructures and Jiangsu Physical Science Research Center, Nanjing University, Nanjing 210093, China}

\author{Hao Xu}
\thanks{These authors contributed equally to this work.}
\affiliation{National Laboratory of Solid State Microstructures and School of Physics, Nanjing University, Nanjing 210093, China} %

\author{Weiqin Zhu}
\thanks{These authors contributed equally to this work.}
\affiliation{Key Laboratory of Computational Physical Sciences (Ministry of Education), Institute of Computational Physical Sciences, State Key Laboratory of Surface Physics, and Department of Physics, Fudan University, Shanghai 200433, China} %

\author{Shufan Cheng}
\affiliation{National Laboratory of Solid State Microstructures and School of Physics, Nanjing University, Nanjing 210093, China} %

\author{Yanyan Shangguan}
\affiliation{National Laboratory of Solid State Microstructures and School of Physics, Nanjing University, Nanjing 210093, China} %

\author{Song Bao}
\affiliation{National Laboratory of Solid State Microstructures and School of Physics, Nanjing University, Nanjing 210093, China} %

\author{Junbo Liao}
\affiliation{National Laboratory of Solid State Microstructures and School of Physics, Nanjing University, Nanjing 210093, China} %

\author{Bo Zhang}
\affiliation{National Laboratory of Solid State Microstructures and School of Physics, Nanjing University, Nanjing 210093, China} %

\author{Zihang Song}
\affiliation{National Laboratory of Solid State Microstructures and School of Physics, Nanjing University, Nanjing 210093, China} %

\author{Shuai Dong}
\affiliation{National Laboratory of Solid State Microstructures and School of Physics, Nanjing University, Nanjing 210093, China} %

\author{Maofeng Wu}
\affiliation{National Laboratory of Solid State Microstructures and School of Physics, Nanjing University, Nanjing 210093, China} %

\author{Stanislav E. Nikitin}
\affiliation{PSI Center for Neutron and Muon Sciences, Paul Scherrer Institute, 5232 Villigen PSI, Switzerland}

\author{Travis J. Williams}
\affiliation{ISIS Neutron and Muon Source, Rutherford Appleton Laboratory, Didcot, OX11 0QX, UK}

\author{Changsong Xu}
\email{csxu@fudan.edu.cn}
\affiliation{Key Laboratory of Computational Physical Sciences (Ministry of Education), Institute of Computational Physical Sciences, State Key Laboratory of Surface Physics, and Department of Physics, Fudan University, Shanghai 200433, China} %

\author{Jinsheng Wen}
\email{jwen@nju.edu.cn}
\affiliation{National Laboratory of Solid State Microstructures and School of Physics, Nanjing University, Nanjing 210093, China} %
\affiliation{Collaborative Innovation Center of Advanced Microstructures and Jiangsu Physical Science Research Center, Nanjing University, Nanjing 210093, China}
\affiliation{Jiangsu Key Laboratory of Quantum Information Science and Technology, Nanjing University, Suzhou, 215163, China}

\date{\today}
\begin{abstract}
	The bond-dependent anisotropic Kitaev interactions are the key for the Kitaev model, which has attracted intense interest for its potential to host quantum-spin-liquid states and fractional excitations. However, experimental realizations of such interactions remain scarce. Here, we investigate the magnetic excitations of \ce{NiI2}, a van der Waals magnet with spin $S=1$. By combining inelastic neutron scattering, magnetization measurements, magnetic structure analysis, first-principles calculations, and linear-spin-wave simulations, we identify a minimal model that features substantial Kitaev and off-diagonal $\Gamma$ interactions, which together stabilize the canted magnetic ground state and open a gap in the spin-wave spectrum. Notably, these interactions arise from strong spin-orbit coupling on the ligand ions, despite the quenched orbital moment of the magnetic \ce{Ni^{2+}} ions. Our results provide compelling experimental evidence for the ligand-driven Kitaev mechanism. This demonstrates a concrete pathway to generating strong bond-dependent anisotropy in systems where the magnetic ions themselves have weak spin-orbit coupling, thereby substantially broadening the range of potential Kitaev materials.
\end{abstract}

\maketitle



Magnetic systems with bond-dependent anisotropic interactions have attracted increasing attention due to their potential to realize exotic quantum phases. The seminal spin-1/2 honeycomb Kitaev model demonstrates how such anisotropic Kitaev interactions can give rise to a quantum-spin-liquid (QSL) ground state and anyonic excitations~\cite{kitaevAnyonsExactlySolved2006}. Realizing this model in real materials is of great interest, both for advancing our understanding of quantum magnetism and for potential applications in topological quantum computation~\cite{broholmQuantumSpinLiquids2020,chouQuantumSpinLiquid2025,kitaevFaulttolerantQuantumComputation2003a}. However, for a spin-only system, it is unrealistic to have such bond-dependent interactions as the three neareast-neighboring bonds of the honeycomb lattice are structurally equivalent.
A major breakthrough came with the proposal that spin-1/2 magnetic ions with strong spin-orbit coupling (SOC), embedded in edge-shared ligand octahedra, can host dominant Kitaev interactions~\cite{jackeliMottInsulatorsStrong2009a, takagiConceptRealizationKitaev2019}. The SOC of the magnetic ion entangles spin and orbital degrees of freedom to form an effective spin $J_{\rm eff} = 1/2$. Thanks to the spatially anisotropic orbitals, bond-dependent exchange anisotropy may be realized. A representative example is $\alpha$-\ce{RuCl3}, a layered honeycomb-lattice compound with $J_{\rm eff} = 1/2$, where the large anisotropic interactions (especially the Kitaev term) are believed to be responsible for various interesting phenomena, including the damped magnon dispersions~\cite{winterBreakdownMagnonsStrongly2017,ranSpinWaveExcitationsEvidencing2017,maksimovRethinkingAlphaRuCl2020a}, fractional excitations~\cite{banerjeeNeutronScatteringProximate2017,doMajoranaFermionsKitaev2017,yokoiHalfintegerQuantizedAnomalous2021}, and QSL state under magnetic fields~\cite{zhouPossibleIntermediateQuantum2023,czajkaOscillationsThermalConductivity2021,yokoiHalfintegerQuantizedAnomalous2021}.
Recent developments have expanded the scope of Kitaev physics to a broader class of materials, which includes systems with different geometries such as triangular~\cite{kimchiKitaevHeisenbergModelsIridates2014,jackeliQuantumOrderDisorder2015,wangComprehensiveStudyGlobal2021,kimBonddependentAnisotropyMagnon2023} and kagome lattices~\cite{kimchiKitaevHeisenbergModelsIridates2014,ghazanfariTopologicalPhaseTransition2017,moritaGroundstatePhaseDiagram2018}, as well as systems with high-spin moments~\cite{stavropoulosMicroscopicMechanismHigherSpin2019,guSignaturesKitaevInteractions2024,shangguanOnethirdMagnetizationPlateau2023,kogaGroundstateThermodynamicProperties2018,zhuMagneticFieldInduced2020,leeTensorNetworkWave2020,khaitCharacterizingSpinoneKitaev2021,xuInterplayKitaevInteraction2018,xuPossibleKitaevQuantum2020,pengKitaevPhysicsTwodimensional2024}.
Among the high-spin Kitaev materials, of particular interest are recent theoretical proposals suggesting that in certain \ce{Ni}-based compounds with edge-shared octahedral coordination~\cite{stavropoulosMicroscopicMechanismHigherSpin2019,pengKitaevPhysicsTwodimensional2024}, conventional Heisenberg exchange may be suppressed, while sizable Kitaev interactions can emerge from strong SOC of the ligand ions and strong Hund's coupling between the two electrons occupying the Ni$^{2+}$ $e_g$ orbitals. Notably, the \ce{Ni^{2+}} ion itself has quenched orbital angular momentum in an octahedral crystal field, leading to a pure $S=1$ spin triplet ground state without spin-orbit entanglement~\cite{abragamElectronParamagneticResonance1970}.
Therefore, these predictions suggest a new route for realizing Kitaev physics---purely through ligand SOC rather than SOC of the magnetic ion. Experimental validation of this mechanism is therefore of critical importance. In this context, \ce{NiI2}, a simple binary compound composed solely of Ni$^{2+}$ ions and iodine ions with strong SOC, and characterized by an edge-shared octahedral geometry, emerges as an excellent candidate.

\begin{figure}[htbp]
	\includegraphics[width=8.6cm]{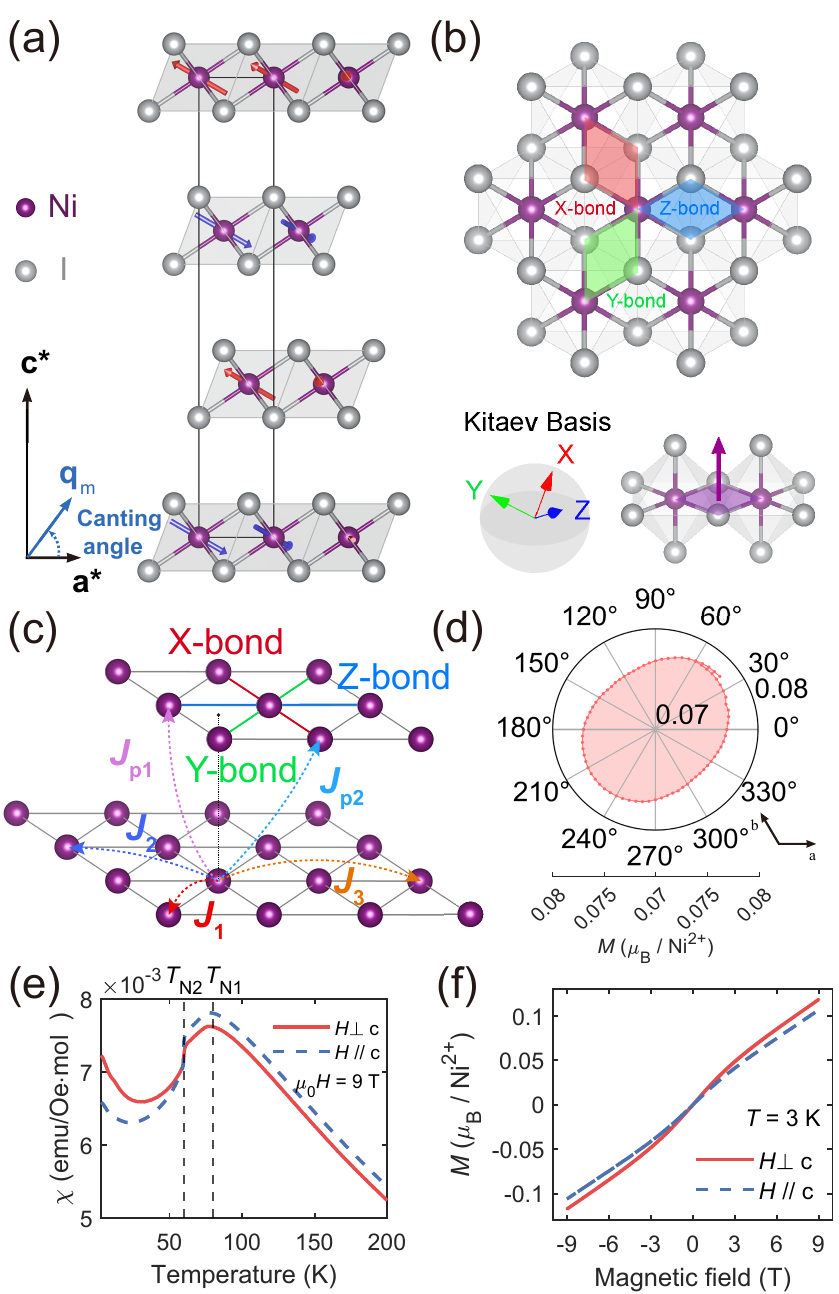}
	\caption{\label{fig: 1} Magnetic structure, Kitaev geometry, and magnetization of single-crystalline \ce{NiI2}.
		(a) Schematic magnetic structure with propagation vector $\vb{q}_{\rm m}$ and canting angle $\theta$.
		(b) Kitaev geometry in a single layer and the Kitaev basis. The normal of a representative Ni-I-Ni bond plane (rhombus) defines the type of the bond.
		(c) Exchange paths and corresponding $X$, $Y$, and $Z$ bonds.
		(d) Angle-dependent magnetization $M$ at $T=2$~K under a 6~T field rotated in the $a$-$b$ plane.
		(e) Magnetic susceptibility $\chi(T)$ measured under $\mu_0 H=9$~T applied parallel (dashed) and perpendicular (solid) to $c$; vertical lines mark $T_{\rm N1}\approx80$~K and $T_{\rm N2}=60$~K.
		(f) Field dependence of $M$ up to 9~T with $\mu_0 H \parallel$ (dashed) and $\perp$ (solid) $c$.}
\end{figure}

\ce{NiI2} is a layered van der Waals magnet that crystallizes in the centrosymmetric \ce{CdCl2}-type structure (space group $R\bar{3}m$, No.166). Within each layer, magnetic \ce{Ni^{2+}} ions ($S = 1$) form a triangular lattice through edge-shared \ce{I^{-}} octahedra [Fig.~\ref{fig: 1}(a)-(c)]. Magnetization measurements, as shown in Fig.~\ref{fig: 1}(e), reveal two successive magnetic transitions in the bulk: a collinear antiferromagnetic order sets in at $T_{\rm N1} \approx 80$~K, followed by a first-order transition at $T_{\rm N2} = 60$~K into a canted proper screw (PS) state~\cite{kuindersmaMagneticStructuralInvestigations1981}.
Across the first-order transition, the lattice undergoes a minute distortion~\cite{kuindersmaMagneticStructuralInvestigations1981}; nevertheless, owing to its small magnitude, all reciprocal lattice vectors are indexed using the high-temperature hexagonal unit cell throughout this work.
In the low-temperature PS phase, spins rotate within a plane perpendicular to the magnetic propagation vector $\vb{q}_{\rm m} = (0.138,\,0,\,1.457)$~\cite{kuindersmaMagneticStructuralInvestigations1981}. This noncollinear magnetic structure breaks inversion symmetry and retains only a two-fold rotation axis combined with time reversal. Such a symmetry breaking gives rise to an in-plane ferroelectric polarization, experimentally confirmed in bulk samples~\cite{kurumajiMagnetoelectricResponsesInduced2013}, establishing \ce{NiI2} as a type-II multiferroic.
Such an intriguing magnetic ground state has attracted significant research interest in \ce{NiI2}~\cite{songElectricalSwitchingPwave2025,aminiAtomicscaleVisualizationMultiferroicity2024,antaoElectricFieldControl2024,liuDensityFunctionalTheory2023,songEvidenceSinglelayerVan2022,jiangDilemmaOpticalIdentification2023,wuLayerThicknessCrossover2023,wangOrientationselectiveSpinpolarizedEdge2024,miaoSpinresolvedImagingAtomicscale2025,aminiAtomicscaleVisualizationMultiferroicity2024,liuCompetingMultiferroicPhases2024,liuSurprisingPressureinducedMagnetic2025,wangMicroscopicEvidenceSpindriven2026}.

Recent theoretical studies suggest that conventional Heisenberg interactions are insufficient to account for the observed ground state, highlighting the potential important role of bond-dependent Kitaev-type interactions~\cite{congSoftMagnonsVan2024,liRealisticSpinModel2023}. This aligns with earlier theoretical predictions proposing the presence of Kitaev interactions in \ce{NiI2}~\cite{stavropoulosMicroscopicMechanismHigherSpin2019}. To rigorously test these hypotheses, an accurate determination of the exchange interactions is highly desirable.
In this study, we present detailed inelastic neutron scattering (INS) measurements and theoretical modeling, leading to the construction of an effective spin Hamiltonian that captures the essential physics of the magnetic ground state as well as excitations, and provides crucial insights into the origin of anisotropic exchange in \ce{NiI2}.


\begin{figure*}[t]
	\includegraphics[width=17.5cm]{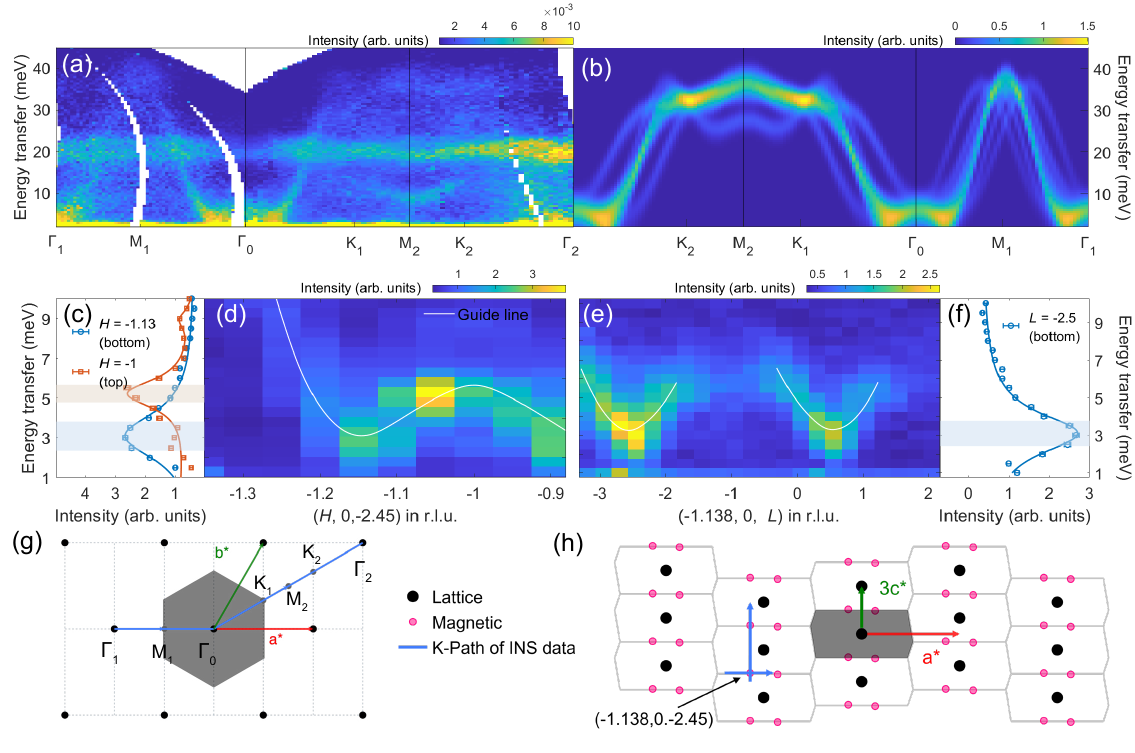}
	\caption{\label{fig: 2} INS data and LSWT calculations in the canted PS phase of \ce{NiI2}.
		(a) Symmetrized INS intensity along the path shown in (g) measured at $T=5$~K. Because of the weak dispersion along $L$ as shown in (e), the data are integrated over $L\in[-1,1]$. Features near 20~meV arise from phonons and Al background.
		(b) Corresponding LSWT dispersions, averaged over six magnetic domains and integrated over $L\in[-1,1]$.
		(c) Const-$Q$ cut at $H=-1.13$ showing a 3.1~meV gap and a kink at 5.2~meV.
		(d),(e) Low-energy dispersions near $(-1.138,0,-2.45)$ along $H$ and $L$ at $T=1.7$~K. White lines are guides to the eye.
		(f) Const-$Q$ cut at $L=-2.5$ confirming the 3.1~meV gap.
		Error bars denote $\sqrt{N}$ counting statistics.
		(g) and (h) are 2D reciprocal lattice in the $(HK0)$ plane (g) and $(H0L)$ plane (h). Black points, pink points and blue paths denote the Lattice Bragg peaks, magnetic Bragg peaks and the paths of INS data shown in (a), (d) and (e). }
\end{figure*}

We investigated the exchange interactions in \ce{NiI2} using inelastic neutron scattering on single crystals [see Supplemental Material (SM)~\cite{SeeSupplementalMaterial} for experimental details]. The in-plane magnon dispersion measured on MAPS [Fig.~\ref{fig: 2}(a)], integrated over $L\in[-1,1]$ due to weak interlayer dispersion [Fig.~\ref{fig: 2}(e)], extends from the Brillouin-zone center to 37.5~meV at the zone boundary and exhibits a nearly flat band top along $\Gamma_0 - M_2 -\Gamma_2$. Low-energy measurements on EIGER [Fig.~\ref{fig: 2}(c)-(f)] reveal a 3.1~meV gap at $(-1.138,0,-2.45)$ and a 5.2~meV kink forming a Mexican-hat dispersion. Along $L$, the spectral weight near 6~meV at the zone boundary is strongly suppressed, while asymmetric group velocities indicate a dominant second-nearest-neighbor (NN) interlayer coupling $J_{\rm p2}$ (see SM~\cite{SeeSupplementalMaterial}). We note that INS measurements on \ce{NiI2} were reported in Ref.~\cite{kimKitaevInteractionProximate2026}; however, the band top and low-energy features including the gap were not resolved due to the limited energy range covered (up to 20~meV) and limited resolution, thereby hindering an accurate determination of the spin Hamiltonian.

With these spectral features in hand, we proceed to construct a minimal magnetic Hamiltonian for \ce{NiI2}.
The presence of a 3.1~meV gap, clearly resolved in Fig.~\ref{fig: 2}(d) and (e), necessitates the inclusion of magnetic anisotropy. Symmetry allows two classes of anisotropic terms: a single-ion anisotropy (SIA) $A_{zz}(S^z)^2$ and bond-dependent anisotropic exchanges, {\it i.e.}, Kitaev interaction $K$, and off-diagonal interactions $\Gamma$ and $\Gamma'$.
The magnetic susceptibility data shown in Fig.~\ref{fig: 1}(d) and (f) indicate a weak in-plane easy-axis anisotropy. The observed anisotropy, however, cannot be explained by SIA alone, as the only symmetry-allowed easy axis is the $z$-axis. This indicates that the origin of the anisotropy must be dominated by bond-dependent anisotropic interactions. By calculating the spin-wave spectra using linear spin-wave theory (LSWT) as implemented in the SpinW package {\new which employs a rotating-frame method for incommensurate structures}~\cite{tothLinearSpinWave2015}, we find that the $\Gamma$ term can give rise to this gap. Moreover, a Kitaev interaction is also necessary, as it is responsible for the finite canting angle between the spin precession plane and $c$-axis, as will be discussed later.
In addition, isotropic Heisenberg exchanges are included up to third-NN within the plane, as the incommensurate magnetic order originates from the competition between ferromagnetic NN exchange $J_1$ and antiferromagnetic third-NN exchange $J_3$~\cite{liRealisticSpinModel2023}. For the interlayer coupling, only $J_{\rm p2}$ is retained, consistent with first-principles calculations and our neutron scattering results.

Guided by symmetry analysis, magnetic anisotropy measurements and insights from first-principles studies of \ce{NiI2}~\cite{liRealisticSpinModel2023,stavropoulosMicroscopicMechanismHigherSpin2019}, we consider the following minimal model (see SM~\cite{SeeSupplementalMaterial} for further details about the model):
\begin{equation}
	\begin{aligned}
		H = & \sum_{\expval{i, j} \in \alpha \beta (\gamma) } J_1 \vb{S}_i \cdot \vb{S}_j + K S_i^\gamma S_j^\gamma  + \Gamma (S_i^\alpha S_j^\beta +S_i^\beta S_j^\alpha  ) \\
		+   & \sum_{\expval{\expval{i, j}}} J_2 \vb{S}_i \cdot \vb{S}_j
		+ \sum_{\expval{\expval{\expval{i, j}}}} J_3 \vb{S}_i \cdot \vb{S}_j
		+ \sum_{\expval{\expval{i, j}}^\perp} J_{\rm p2} \vb{S}_i \cdot \vb{S}_j
	\end{aligned}
	\label{Eq: minimal model}
\end{equation}
where $\expval{i, j}$, $\expval{\expval{i, j}}$ and $\expval{\expval{\expval{i, j}}}$ denote pairs of first, second and third NN within layers, while $\expval{\expval{i, j}}^{\perp }$ refer to second NN interlayer coupling; $\alpha,\beta,\gamma$ lable directions in the Kitaev basis \{$\hat{X}, \hat{Y}, \hat{Z}$\}, with $\gamma$ along the direction indicated by the bond. The Kitaev basis, corresponding bonds, and interactions are illustrated in Fig.~\ref{fig: 1}(b) and (c).

\begin{figure}[htbp]
	\includegraphics[width=8.6cm]{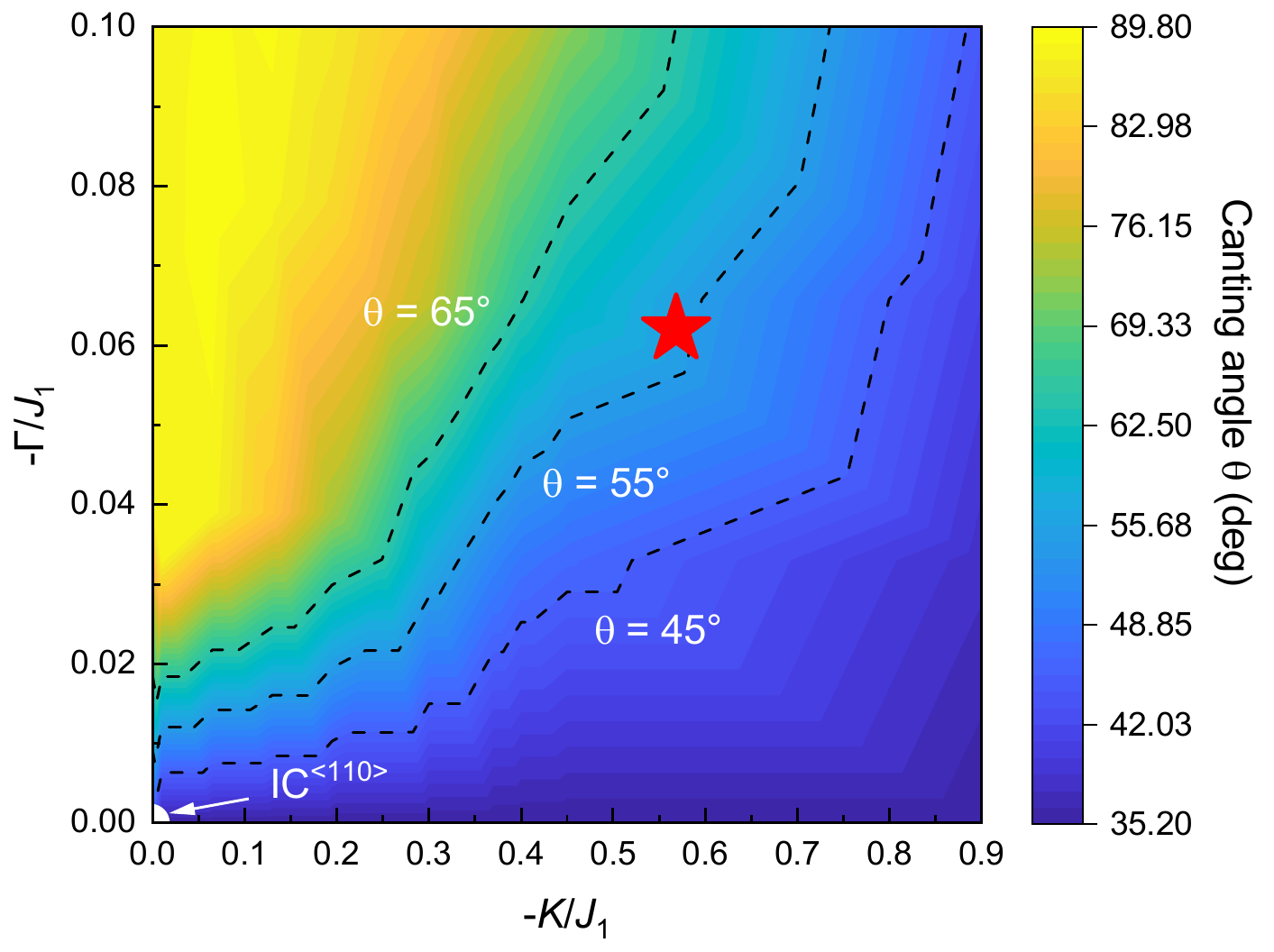}
	\caption{\label{fig: phase diagram}
		Phase diagram of Eq.~\ref{Eq: minimal model}.
		Parameters are fixed to $J_1=-1$, $J_2=0.086$, $J_3=0.33$, and $J_{\rm p2}=0.14$.
		The IC$^{\langle110\rangle}$ phase appears only at $K=\Gamma=0$ (white region), while finite $K$ or $\Gamma$ stabilizes the canted PS$^{\langle1\bar{1}0\rangle}$ phase, consistent with Ref.~\cite{liRealisticSpinModel2023}.
		The color scale shows the canting angle; contours at $45^\circ$, $55^\circ$, and $65^\circ$ are indicated, and the red star marks the optimal parameters.
	}
\end{figure}

As a successful minimal model, we find a clear one-to-one correspondence between the features observed in the experiment and specific exchange parameters. For instance, the 3.1~meV gap originates from a small AFM $\Gamma$ term, the 5.2~meV kink results from an AFM $J_3$, and the flat band top along $\Gamma_0-M_2-\Gamma_2$ path is due to the presence of a finite AFM $J_2$.
On the other hand, we find that accurately determining the Kitaev term $K$ requires taking into account not only the INS spectra but also the peculiar magnetic structure. In particular, the canting angle of the spin rotation plane — reported as $55^\circ \pm 10^\circ$~\cite{kuindersmaMagneticStructuralInvestigations1981} and illustrated in Fig.~\ref{fig: 1}(a) — depends sensitively on $K$.
Classical Monte Carlo simulations (Fig.~\ref{fig: phase diagram}) show that AFM $K$ and $\Gamma$ stabilize the canted PS state along crystal orientation $[1\bar{1}0]$, but with opposite effects on spin orientation: the Kitaev term drives the spins out of the $a$-$b$ plane, whereas the $\Gamma$ term favors their rotation within the plane.
This contrasting tendency indicates that the eventual canting angle should emerge from the delicate balance between the two competing contributions---our analysis shows that $K/\Gamma \approx 9$ will give a proper canting angle.
By jointly and self-consistently constraining the spin-wave spectra and the magnetic structure—specifically the canting angle, in-plane propagation direction, and period—we iteratively refine the model and obtain the following optimal parameters:
$J_1= -5.86(0.122),\, K = 3.33(0.032),\, \Gamma = 0.37(0.007), \, J_2= 0.28 (0.089), \, J_3= 1.99 (0.024), \, $ and $J_{\rm p2} = 0.78(0.005)$.
The parameters are given in meV, with values in parentheses indicating their uncertainties, yielding a reduced $\chi^2 = 0.637$.
Under these parameters, the magnetic structure stabalizes into the canted PS state with a canting angle of 53.4$^\circ$ and an in-plane propagation along crystal orientation $[1\bar{1}0]$, modulated with a period of 7$a$.
Both values are in excellent agreement with the experimental results, which report a canting angle of $55^\circ$ and a period of $7.23a$~\cite{kuindersmaMagneticStructuralInvestigations1981}.
Compared to Ref.~\cite{kimKitaevInteractionProximate2026}, the stronger constraints imposed by the band top and bottom, as well as magnetic structure in our work result in a smaller $J_1$, a larger Kitaev interaction, and a finite off-diagonal $\Gamma$ term.

In Fig.~\ref{fig: 2}(b), Figs.~S3-S4 in SM~\cite{SeeSupplementalMaterial}, we present the calculated spectra based on LSWT dispersions. To enable a direct comparison with the experiment, all spectra are averaged over six magnetic domains characterized by the propagation vectors $\pm (0.138,0,1.457) $, as well as four additional vectors generated by $120^\circ$ and $240^\circ$ rotations about the $c$-axis.
Figure~\ref{fig: 2}(b) reproduces the overall dispersion well. However, the high-energy spectral intensity—particularly near the $K$ point—is significantly weaker in the experimental data than in the LSWT calculation. We attribute this discrepancy to two main factors.
First, the noncollinear magnetic structure and sizable Kitaev interaction in \ce{NiI2} can give rise to strong anharmonic effects, leading to magnon decay, spectral broadening, and reduced spectral weight, as observed in related systems~\cite{kimBonddependentAnisotropyMagnon2023,xieQuantumSpinDynamics2024}. Consistently, the calculated two-magnon density of states (Fig.~S6 in SM~\cite{SeeSupplementalMaterial}) exhibits pronounced intensity near the $K$ and $M$ points, supporting this interpretation.
Second, magnon Umklapp scattering processes are not included in the SpinW calculations~\cite{tothLinearSpinWave2015}. Although their contribution is expected to be subleading in a high-spin system, their omission may also contribute to the overestimated spectral weight near the band top.
As shown in Fig.~S4(a), the calculated spectra also reproduce the apparent magnon intensity weakening observed at the band top along the $L$ direction in Fig.~\ref{fig: 2}(e). We also performed constant-energy cuts from the band bottom to the top and computed the corresponding intensity patterns. As shown in Fig.~S3 in SM~\cite{SeeSupplementalMaterial}, the magnon excitations progressively expand from the Brillouin zone center toward the zone boundary with increasing energy. Due to bond-dependent anisotropy and the presence of multiple magnetic domains, the outer boundary of the excitations takes on a characteristic hexagonal shape.
The agreement between experiment and theory further supports the validity of the our model parameters.

{\new 
It is worth noting that the gap observed in our experiment (approximately 3~meV) differs from the sub-0.3~meV gap reported in Ref.~\cite{kimKitaevInteractionProximate2026}. Based on our model, we believe that this discrepancy originates from both the different magnon branches being probed and the different criteria used to define the gap.
In the incommensurate magnetic state, each Ni ion in a unit cell gives rise to three magnon branches, namely $\omega(\mathbf{k})$, $\omega(\mathbf{k}+\mathbf{q}_{\rm m})$, and $\omega(\mathbf{k}-\mathbf{q}_{\rm m})$. Among them, the $\omega(\mathbf{k})$ branch carries the dominant spectral weight, whereas the two satellite branches possess substantially weaker intensity, despite having somewhat lower band-bottom energies. These satellite branches can indeed be identified in our calculations [Fig.~\ref{fig: 2}(b)], where their spectral weight is found to be extremely weak.
In our triple-axis neutron measurements, the observed gap was associated with the dominant $\omega(\mathbf{k})$ branch, whose dispersion provided the primary constraints on the determination of the model parameters. By contrast, in Ref.~\cite{kimKitaevInteractionProximate2026}, the neutron intensity was integrated over the $(H,K)$ plane and the analysis focused on the low-energy region below 2~meV. Under such conditions, the weak spectral-weight contributions from the $\omega(\mathbf{k}\pm\mathbf{q}_{\rm m})$ branches become visible and can fill in the low-energy gap.
Furthermore, the gap definitions adopted in the two studies were not identical. In our work, the gap was determined from the peak position of a Lorentzian fit to constant-$Q$ cuts, as shown in Fig.~\ref{fig: 2}(c) and (f). In contrast, Ref.~\cite{kimKitaevInteractionProximate2026} estimated the gap from the energy below which the magnetic scattering intensity became nearly indistinguishable from the background. Because the magnon excitations in \ce{NiI2} exhibit substantial intrinsic broadening, this latter criterion naturally tends to yield a smaller gap.}

\begin{figure}[htbp]
	\includegraphics[width=8.6cm]{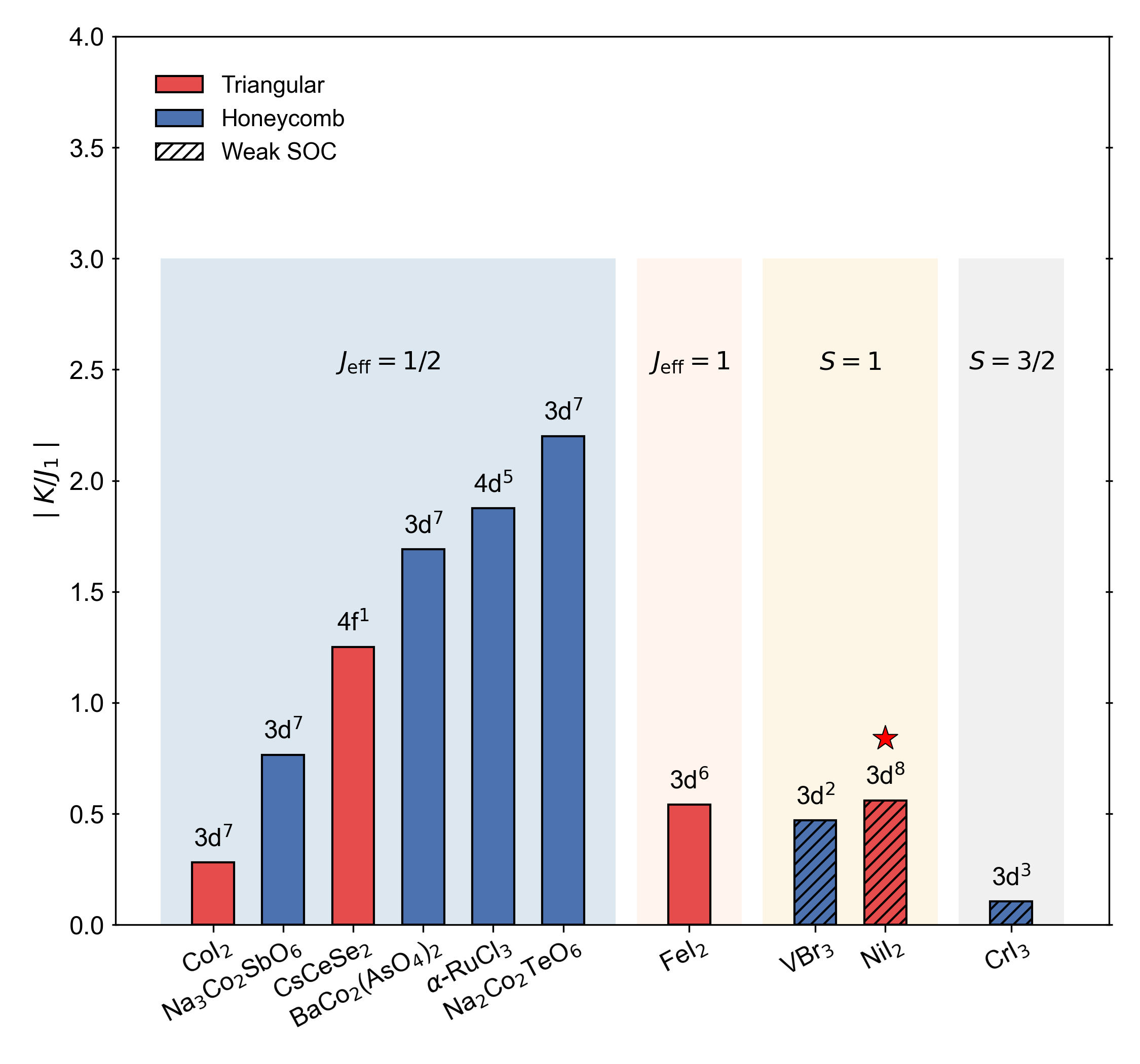}
	\caption{\label{fig: Kitaev_sum}
		List of experimentlly identified Kitaev materials.The $y$ axis shows $|K/J_1|$. Red (blue) bars denote triangular (honeycomb) lattices. Hatched bars indicate weak magnetic-ion SOC. Refs.~\cite{kimBonddependentAnisotropyMagnon2023,kimAntiferromagneticKitaevInteraction2022,xieQuantumSpinDynamics2024,maksimovStrongKitaevInteraction2025,maksimovRethinkingAlphaRuCl2020a, baiHybridizedQuadrupolarExcitations2021,kaoKitaevInteractionsVan2026a,kimSpinLatticeDynamics2024}
	}
\end{figure}
In constructing the minimal spin Hamiltonian, we note that, beyond the bilinear interactions considered here, a biquadratic term in the form of $B(\vb{S}_i \cdot \vb{S}_j)^2$ may be present in this system~\cite{liRealisticSpinModel2023,niGiantBiquadraticExchange2021}. The most direct experimental signature of such a term would be the emergence of dispersive multipolar excitations at high energy. However, since neutrons are only weakly sensitive to multipolar excitations, our INS measurements do not reveal clear signatures that would allow a reliable determination of the biquadratic coupling. We therefore do not include this term in the present minimal effective model, which captures the essence of the physics in \ce{NiI2} already. Future investigations using probes more sensitive to multipolar excitations, such as resonant inelastic x-ray scattering, may help clarify its possible role.

So far, we have demonstrated the presence of a sizable Kitaev interaction in \ce{NiI2} where the orbital moment of the magnetic \ce{Ni^{2+}} ions are quenched. In Fig.~\ref{fig: Kitaev_sum}, we plot the values of $|K/J_1|$ for some typical  materials together with \ce{NiI2}.
In $J_{\rm eff}=1/2$ systems~\cite{kimBonddependentAnisotropyMagnon2023,kimAntiferromagneticKitaevInteraction2022,xieQuantumSpinDynamics2024,maksimovStrongKitaevInteraction2025,maksimovRethinkingAlphaRuCl2020a}, the presence of Kitaev interactions have been well established by multiple experimental probes, with $\alpha$-\ce{RuCl3} as the most prominent example~\cite{winterBreakdownMagnonsStrongly2017,ranSpinWaveExcitationsEvidencing2017,maksimovRethinkingAlphaRuCl2020a, banerjeeNeutronScatteringProximate2017,doMajoranaFermionsKitaev2017,yokoiHalfintegerQuantizedAnomalous2021,zhouPossibleIntermediateQuantum2023,czajkaOscillationsThermalConductivity2021,yokoiHalfintegerQuantizedAnomalous2021}.
As research on the Kitaev physics continues to advance, attention is increasingly extending beyond the $S=1/2$ limit to higher-spin systems.
{\new In $J_{\rm eff} = 1$ triangular magnet \ce{FeI2}, the bond-dependent anisotropy leads to hybridization of one-magnon and single-ion bound state~\cite{baiHybridizedQuadrupolarExcitations2021}.
}
For $S=1$ magnets, \ce{VBr3} exhibits a moderate Kitaev interaction~\cite{kaoKitaevInteractionsVan2026a}. For its isostructural ligand-substituted counterpart \ce{VI3}, whether a $K$ term exists remains under debate~\cite{guSignaturesKitaevInteractions2024,shenMagnetoelasticHoneycombFragmentation2026,guptaAngulardependentThermalHall2026}. For the $S=3/2$ material \ce{CrI3}, the presence of a Kitaev interaction is also debated~\cite{kimSpinLatticeDynamics2024,chenMagneticAnisotropyFerromagnetic2020,leeFundamentalSpinInteractions2020,stavropoulosMagneticAnisotropySpin32021}, whereas in \ce{CrBr3}~\cite{caiTopologicalMagnonInsulator2021,nikitinThermalEvolutionDirac2022} and \ce{CrCl3}~\cite{doGapsTopologicalMagnon2022} it is generally considered to be absent.
In \ce{NiI2}, although the Kitaev interaction appears to be smaller than that in some $J_{\rm eff}-1/2$ systems with strong SOC on the magnetic ions, a value of 3.33(0.032)~meV is remarkably large given the quenched orbital moment of the \ce{Ni^{2+}} ions. This provides a promising route to realizing significant Kitaev interactions through the strong SOC of the ligand ions.
Although quantum fluctuations are reduced at higher spin,  bond-dependent anisotropic interactions can still be highly consequential, profoundly influencing the magnetic ground state or excitation spectrum~\cite{shangguanOnethirdMagnetizationPlateau2023,guSignaturesKitaevInteractions2024,kogaGroundstateThermodynamicProperties2018,zhuMagneticFieldInduced2020,leeTensorNetworkWave2020,khaitCharacterizingSpinoneKitaev2021,xuPossibleKitaevQuantum2020,pengKitaevPhysicsTwodimensional2024}. However, investigations of Kitaev interactions in high-spin systems are still in their early stage and therefore require further exploration.

Most existing studies of Kitaev physics have focused on honeycomb lattices, while triangular-lattice systems have received comparatively less attention. As we demonstrate in \ce{NiI2}, a triangular-lattice system can also host bond-dependent $K$ and $\Gamma$ interactions. In fact, in the presence of an edge-shared octahedral ligand environment, bond-dependent interactions can naturally arise in triangular-lattice systems as well~\cite{kimBonddependentAnisotropyMagnon2023,xieQuantumSpinDynamics2024,xieDominantKitaevInteraction2025,ortizQuantumDisorderedGround2023}. This highlights triangular lattice as a promising and largely unexplored platform for Kitaev physics.

In summary, based on the magnetic excitation spectra, together with magnetization, magnetic structure, first-principle calculations and LSWT simulations, we firmly establish the presence of sizable Kitaev interaction in a simple binary compound \ce{NiI2}, where the orbital angular momentum of the magnetic ions is fully quenched, demonstrating that substantial Kitaev interactions can arise purely from the strong SOC of the ligand ions. This finding paves the way for identifying bond-dependent anisotropic interactions and exploring the associated rich physics in a broader class of magnetic systems.

\bigskip

$Acknowledgments$---We thank Zhao-Yang Dong for stimulating discussions. The work was supported by the National Key Projects for Research and Development of China (Grants No.~2024YFA1409200, and No.~2021YFA1400400), the National Natural Science Foundation of China (Grants No.~12225407, No.~12434005 and No.~12404173), the Natural Science Foundation of Jiangsu Province (Grants No.~BK20233001, No.~BK20241251, and No.~BK20241250), the China Postdoctoral Science Foundation (Grants No.~BX20240161 and No.~2024M751367), the Jiangsu Funding Program for Excellent Postdoctoral Talent (Grant No.~2024ZB021), the Xiaomi Young Scholars Technology Innovation Award, and the Fundamental Research Funds for the Central Universities (Grant No.~KG202501). C.X. acknowledges support from the National Natural Science Foundation of China (Grant No.~12274082), Shanghai Pilot Program for Basic Research—FuDan University No.~21TQ1400100 (23TQ017), Shanghai Science and Technology Committee (No.~23ZR1406600), Innovation Program for Quantum Science and Technology (2024ZD0300102), and the Xiaomi Young Talents Program. Part of this work is based on experiments performed at ISIS (Proposal No.~RB2420024, DOI: 10.5286/ISIS.E.RB2410072), and the Swiss spallation neutron source SINQ, Paul Scherrer Institute, Villigen, Switzerland.

\bigskip



%

\end{document}